\documentclass{Interspeech}

\usepackage{cite}

\usepackage{graphicx}
\usepackage{subcaption}
\usepackage{booktabs}

\interspeechcameraready

\title{You Sound a Little Tense: L2 Tailored Clear TTS\\ Using Durational Vowel Properties}

\author[affiliation={1,2,3}]{Paige}{Tuttösí}
\author[affiliation={4}]{H. Henny}{Yeung}
\author[affiliation={4}]{Yue}{Wang}
\author[affiliation={2}]{Jean-Julien}{Aucouturier}
\author[affiliation={1}]{Angelica}{Lim}

\affiliation{School of Computing Science}{Simon Fraser University}{Canada}
\affiliation{SUPMICROTECH, CNRS, institut FEMTO-ST}{Université Marie et Louis Pasteur}{France}
\affiliation{}{Enchanted Tools}{France}
\affiliation{Department of Linguistics}{Simon Fraser University}{France}
\email{ptuttosi@sfu.ca}

\usepackage{comment}

\begin{document}

\maketitle

\begin{abstract}
We present the first text-to-speech (TTS) system tailored to second language (L2) speakers. We use duration differences between American English tense (longer) and lax (shorter) vowels to create a ``clarity mode'' for Matcha-TTS. Our perception studies showed that French-L1, English-L2 listeners the participants had fewer (at least 9.15\%) transcription errors when using our clarity mode, and found it more encouraging and respectful than overall slowed down speech. Remarkably, listeners were not aware of these effects: despite the decreased word error rate in clarity mode, listeners still believed that slowing all target words was the most intelligible, suggesting that actual intelligibility does not correlate with perceived intelligibility. Additionally, we found that Whisper-ASR did not use the same cues as L2 speakers to differentiate difficult vowels and is not sufficient to assess the intelligibility of TTS systems for these individuals.
\end{abstract}

\section{Introduction}
Imagine your immigrant family arrives in a new country; you speak the language of your new country at a basic level, but all of a sudden, you need to understand public announcements and alerts on transit, or you need to call to set up your medical card and pass through an automated voice messaging system.  In all of these cases, you will encounter the use of voice-based technology, in recognition, synthesis, and end-to-end systems, which is an area that is currently seeing expansive growth. While these text-to-speech (TTS) systems can help improve accessibility, such as in-home automation systems for the physically and visually impaired \cite{VIEIRA2022121961, deoliviera}, or as a socially supportive device to allow elderly individuals to maintain their autonomy \cite{london}, understanding TTS can be daunting for second-language (L2) speakers when synthesized voices are fast-speaking and directed toward first-language (L1) speakers. 

There remains much work to ensure the ethical validity of such systems in terms of transparency, bias, fairness, and many other factors \cite{ethicsreview}. One aspect of fairness lacking in systems with synthesized speech is their inability to adapt to those with different hearing and comprehension abilities, which is the case for L2 speakers. Previous work in clear TTS has principally explored speech-in-noise \cite{10.1145/1228716.1228732, 10.1007/978-3-030-98358-1_43, novitasari21_interspeech, cohn20_interspeech, 8343873}, rather than factors related to speaker comprehension, like language ability. In human-to-human interaction, the ability of a speaker to adapt to an interlocutor is invaluable: humans modify their speech to L2 speakers \cite{Redmon20}. Yet, evidence shows that human adaptation to L2 speakers does not consistently aid the L2 speakers in their comprehension \cite{AOKI2024101328,ROTHERMICH201922}, therefore training on L2-directed speech may not be the ideal approach. To improve the fairness of voice technology and move towards adaptive synthesized speech, we must first understand how speech can be adjusted to facilitate comprehension in a data-driven, perception-based manner.

To improve TTS for second language speakers, we present a clarity mode for English TTS, built upon a perception-based approach. L2 speakers of English, particularly those whose L1 does not have vowel tensity contrast, often struggle to differentiate English lax (less extreme articulation and shorter) vowels with tense (more extreme articulation and longer) vowels \cite{MonnotMichel1974LPdf, StrangeWinifred2007Avwa, MillerJoanneL.2011DEiS, iverson12, 2019japanese, LU2021101049}. For example, French speakers may replace the lax \textipa{/I/} (ship) by the tense \textipa{/i/} (sheep) \cite{iverson12}. In our paper \cite{tuttosi24_interspeech}, psychophysical reverse-correlation \cite{AHU71} was used to reconstruct L1-English and L2-English listener’s mental representation of duration and pitch in tense-lax vowel contrasts in a data-driven manner. For L2-English speakers who spoke French, Mandarin and Japanese as an L1, it was found that vowel duration, but not pitch, affected perception of tense/lax vowels. This suggests that we can apply a duration mechanism to improve comprehension of this English vowel contrast when an L2 listener struggles to use the primary formant cues \cite{kewley2005influence}. 
Following these results, we create, for the first time, an L2 ``clarity" mode within a TTS to improve L2 comprehension of these difficult American English vowels.

\section{Mechanism validation}
In \cite{tuttosi24_interspeech} we showed that shortening a word with an ambiguous vowel sound resulted in an English-L2 listener hearing the lax vowel, and lengthening the word resulted in hearing the tense vowel. We also explored whether these duration cues could be used to control the perception of \emph{non-ambiguous} vowels in English-L2 \cite{inprep}. Overall, we observed a bias towards the lax vowel, yet English-L2 speakers (with French, Chinese, and Japanese L1s) could be convinced to hear the tense vowel if the duration of a word containing a lax vowel became too long \cite{inprep}. These results suggested that the usual approach for L2-directed speech, slowing either the entire phrase or difficult words for emphasis, can result in lax vowels being mistaken as tense vowels by L2 speakers from a variety of L1 backgrounds, as also shown in clear speech \cite{redmon2020cross}. Instead, to improve L2 comprehension, the duration properties of tense/lax vowels needed to be maintained, applying the emphasis only to tense vowels.

\section{L2 clarity TTS}

Given the improved perception of tense/lax vowels with simple, linguistically-driven duration changes, we added this as a new `clarity mode' in Matcha-TTS \cite{matcha} for L2 speech. 

To enable L2 clarity mode in Matcha-TTS, we added a clarity flag that can be set to ``True'' or ``False'' at synthesis, along with a markup to control which words are emphasized. The user (or large language model driving a dialogue system) surrounds difficult words with exclamation points, e.g., ``!peel!'', allowing the TTS to parse the words to be treated for clarity through several steps: 1) Parse each flagged word to see if it contains a tense or lax vowel, 2) If the word contains tense vowels but no lax vowels, the clarity modification is applied to the tense vowel containing word, 3) If the word contains both tense and lax vowels, and if the tense vowel has primary stress, the clarity modification is applied to the tense vowel containing word.

The modifications are made by applying an array ($carray$) with the same length as the phonemized phrase to scale the predicted duration of each phoneme ($w$), a method shown to result in natural and effective duration changes in TTS systems containing a phonemizer \cite{joly23_ssw}. This \emph{clarity duration} multiplier is applied after the base speech rate multiplier, which is an array containing the speech rate provided at synthesis (in our case, 0.75 which in our pilot was found to have a natural, conversational speech rate for English L1 listeners). The predicted durations ($w$) result from the Hadamard product of the text encoder outputs (Equation~\(\ref{eq:step1}\)): $\log w$: the log duration predicted by the duration predictor as in \cite{grad}, and $x_{\text{mask}}$: the mask for the text input indicating which values are valid \cite{NEURIPS2020_5c3b99e8}. In Matcha-TTS this is then scaled to the input duration the $speechrate$ array. The resulting $y_{\text{lengths}}$ and $y_{\text{max\_length}}$ are then used to calculate the attention alignment map for the text encoder.

 \vspace{-6mm}
 \begin{equation}
 \vspace{-2mm}
\begin{aligned}
    w &= e^{\log w} \odot x_{\text{mask}} \\
    w_{\text{ceil}} &= (\lceil w \rceil \odot \text{speechrate}) \odot \text{c\_array} \\
    y_{\text{max\_length}} &= \max\left(\max\left(1, \sum_{i, j} w_{\text{ceil}}[i, j]\right)\right) \\
\end{aligned}
\label{eq:step1}
\end{equation}

A 1.6x stretch is applied across the entire word with a gradual ramp up and down to the base speech rate over the 6 phonemized items (if there are 6 phonemized items between target words, otherwise as many as are available) preceding and following the target word. Six items were chosen as this encompasses two phonemes preceding and following the target word (approximately 200-300ms \cite{ma21_interspeech} as per \cite{tuttosi24_interspeech}).We also apply a 1x stretch (the base speech rate) to lax-vowel-containing words simultaneously, following the same parsing above to ensure that lax vowels are not stretched by surrounding tense vowel words. A 1.6x stretch was chosen to minimize duration changes. In our validation \cite{inprep}, we saw that perception performance quickly improved from the baseline by increasing the duration of the target, tense vowel-containing word. There was minimal improvement in performance between a 1.6x and 2.0x stretch.

\subsection{Stimulus generation} \label{gen}
We tested 4 different TTS styles: 1. \textbf{Base}: the base Matcha-TTS (0.75x speech rate), 2. \textbf{Stretch}: 1.2x ($0.75\times1.6)$ speech rate applied across the entire phrase, 3. \textbf{Emphasis}: 1.6x stretch applied across all target words (0.75x speech rate elsewhere), and 4. \textbf{Clarity}: 1.6x stretch applied across the target words containing a tense vowel (0.75x speech rate elsewhere).

Previously, in \cite{tuttosi24_interspeech, inprep}, only a \textbf{single target word}, always at the end of a phrase, was tested. To test the robustness of the L2 clarity TTS when the target words occurred in other parts of the phrase, we tested clarity mode in several contexts. First, single target word phrases where the target word was in the middle of the phrase were tested. We then tested \textbf{double target word} phrases (e.g., ``Write down dull and doll''), where the targets could be at the end of the phrase, in the middle of the phrase, or at the beginning of the phrase. Additionally, we explored the performance when the TTS had a combination of tense and lax vowels in a single phrase; we tested a tense and a lax minimal pair, a tense and a lax that are not minimal pairs, two tense and two lax vowels. We aimed to use a variety of starting and ending consonants surrounding the target vowels, and all words had a minimal pair in English. We tested all tense/lax vowel pairs: \textipa{/i/} (peel) and \textipa{/I/} (pill), \textipa{/u/} (fool) and \textipa{/U/} (full),  \textipa{/A/} (cot) and \textipa{/2/} (cut). Lastly, we aimed not to semantically bias phrase meanings towards any word, that is, given the context outside of the target word neither word in the minimal pair made more logical or semantic sense. To do this, we asked ChatGPT-4o to provide a starting list of neutral phrases with varying lengths. These were then selected and modified to create our final list of 16 phrases and insert our target words (Table \ref{phrase list}).

\begin{table}[]
  \centering
  \caption{List of phrases for the experiment}
  \vspace{-2mm}
  \begin{tabular}{|p{7.7cm}|} 
    \hline
    \textbf{Phrases} \\
    \hline
    The word \underline{cut} seemed important to the instructions. \\
    She kept mentioning \underline{cot} during the conversation. \\
    The speaker mentioned \underline{pill}, or at least something similar. \\
    The word \underline{pill} was what she was trying to write. \\
    The phrase had \underline{fool} somewhere in the middle of it. \\
    I saw \underline{full} written on the note pad. \\
    The sign mentioned \underline{sin}, but the person said \underline{scene}. \\
    He wrote down \underline{bought}, but remembered it as \underline{but}. \\
    In his talk he kept using \underline{could}, but I am pretty sure he meant \underline{cooed}. \\
    The paper mentioned \underline{kid}, yet he is telling me \underline{knot}. \\
    There was confusion between \underline{pull} and \underline{bean} in their speech. \\
    I am not sure if the word was \underline{pool} or if \underline{cup} was the right one. \\
    \underline{Sheep} goes on the top of the page and \underline{dull} goes on the bottom. \\
    \underline{Bit} was the first word he said, then \underline{nut} followed. \\
    Actually \underline{hut} is the correct word, it was replaced with \underline{should} by accident. \\
    Maybe he said \underline{hot}, but I really thought \underline{keyed} was what he said. \\
    \underline{Reap} was a more important word in the story than \underline{wooed}. \\
    \hline
    \textbf{Confusion Phrases} \\
    \hline
    The speaker mentioned \underline{pill}, or at least something similar. \\
    He wrote down \underline{but}, but remembered it as \underline{bought}. \\
    There was confusion between \underline{pool} and \underline{bin} in their speech. \\
    Maybe he said \underline{hut}, but I really thought \underline{kid} was what he said. \\
    The word \underline{cot} seemed important to the instructions. \\
    The sign mentioned \underline{scene}, but the person said \underline{sin}. \\
    \underline{Ship} goes on the top of the page and \underline{doll} goes on the bottom. \\
    \underline{Rip} was a more important word in the story than \underline{wood}. \\
    The phrase had \underline{full} somewhere in the middle of it. \\
    In his talk he kept using \underline{cooed}, but I am pretty sure he meant \underline{could}. \\
    I am not sure if the word was \underline{pull} or if \underline{cop} was the right one. \\
    Actually \underline{hot} is the correct word, it was replaced with \underline{shooed} by accident. \\
    She kept mentioning \underline{cut} during the conversation. \\
    The word \underline{peel} was what she was trying to write. \\
    The paper mentioned \underline{keyed}, yet he is telling me \underline{nut}. \\
    \underline{Beat} was the first word he said, then \underline{knot} followed. \\
    I saw \underline{fool} written on the note pad. \\
    \hline
  \end{tabular}
  \label{phrase list}
  \vspace{-8mm}
\end{table}

\begin{figure}[]
  \centering
  \includegraphics[width=1.05\linewidth]{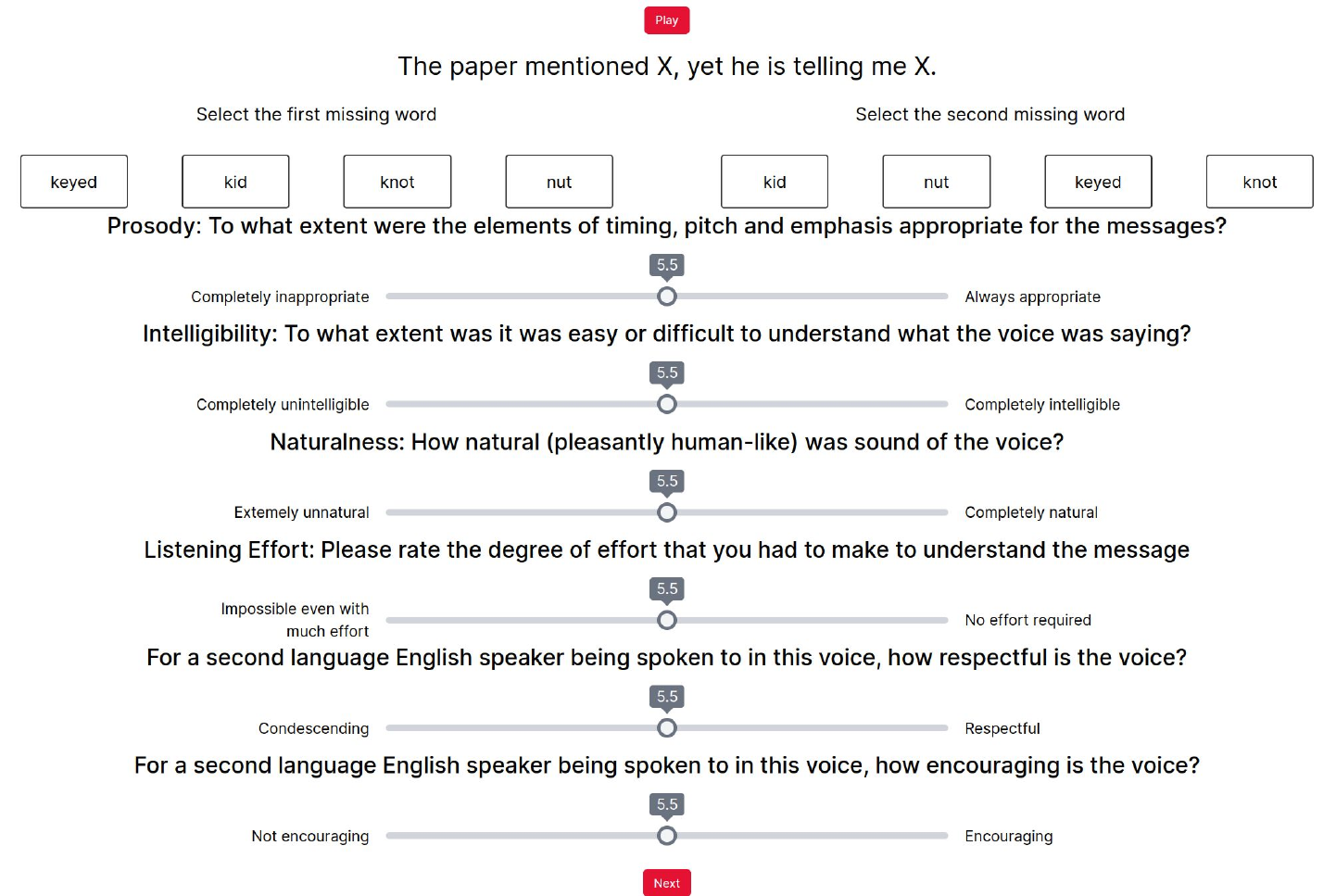}
  \caption{L2-TTS double-word experiment set up in Gorilla.}
  \label{ttsexp}
\vspace{-5mm}
\end{figure}

Because participants would hear each of the phrases 4 times (each TTS style had the same phrases), we added 1 confusion phrase (randomly assigning a TTS style) for each of the phrases. A confusion phrase was the same phrase context using the opposite minimal pair, for example: ``She kept mentioning cot during the conversation.'' and the confusion phrase ``She kept mentioning cut during the conversation.'' Lastly, for the single target word TTS, ``clarity'' TTS was the same as ``emphasis'' for a single tense vowel target and ``base'' for a single lax vowel target. As such, we assessed the TTS in terms of the length of the target words rather than TTS styles and ensured no repeated phrases with the same treatment. 

\begin{table*}[]
  \centering
  \caption{Single target word results: word error rate, intelligibility (iMOS), naturalness (nMOS), effort (eMOS), prosody (pMOS), encouragement (Enc) and respect (Resp.) Our clarity mode, using base lax and target stretch tense, achieves the lowest WER.}
  \vspace{-3mm}
  \begin{tabular}{| l | l | l | l |  l |  l |  l |  l |}
    \hline
    {\textbf{TTS Style}} & {\textbf{WER}} $\downarrow$ & {\textbf{iMOS}} $\uparrow$ & {\textbf{nMOS}} $\uparrow$ & {\textbf{eMOS}} $\uparrow$ & {\textbf{pMOS}} $\uparrow$ & {\textbf{Enc.}} $\uparrow$ & {\textbf{Resp.}} $\uparrow$\\
    \hline
    Base lax (Ours)  & \textbf{16.86}\% & 7.63  & 7.11/*** & 7.33 & 7.20/*** & 6.44/*** & 6.63/** \\
    Target stretch lax & 28.90\% & \textbf{8.18}*/** & \textbf{7.91}**/*** & \textbf{7.73} & \textbf{7.87}*/*** & \textbf{7.03}*/*** & \textbf{7.32}*/***\\
    Full stretch lax &  29.65\% & 7.53 & 4.18 & 7.40 & 5.83 & 5.54 & 5.76\\
    \hline
    Base tense  & 60.23\% & 7.57 & 7.07/***  & 7.18 & 7.22/*** & 6.45/*** & 6.53/** \\
    Target stretch tense (Ours) & \textbf{29.48}\% & \textbf{8.18}*/ & \textbf{7.71}/*** & \textbf{7.69} & \textbf{7.80}/*** & \textbf{6.84}/*** & \textbf{7.00}/***\\
    Full stretch tense &  37.57\% & 7.68 & 4.27 & 7.61 & 6.13 & 5.49 & 5.68\\
    \hline
  \end{tabular}
    \label{single_result}
    {\parbox{6in}{
    \centering
\footnotesize 
Significance values are presented as: Significance compared to base or target stretch / Significance compared to full stretch TTS\\
* = 0.05, ** = 0.01, *** = 0.001
}}
\vspace{-3mm}
\end{table*}

\begin{table*}[]
  \centering
  \caption{Single target word results: ANOVA and Tukey test statistics on MOS Likert scores.}
  \vspace{-3mm}
  \begin{tabular}{|  l |  l |  l  | p{9cm}|}
    \hline
    {\textbf{TTS Style}} & {\textbf{F-Statistic (5,1029)}} & {\textbf{P-Value}} & {\textbf{Significant Tukey Results}}\\
    \hline
    iMOS  &  5.12 & $<$.001 & Lax: targetS/base,fullS p=.043, p=.008; Tense: targetS/base p=.016 \\
    nMOS &  110.83 & $<$.001 & Lax: all/fullS p$<$.001 targetS/base p=.006; Tense: all/fullS p$<$.001 \\
    eMOS & 2.21 & 0.051 & N/A \\
    pMOS & 30.24 & $<$.001 & Lax: all/fullS p$<$.001 targetS/base p=.028; Tense: all/fullS p $<$.001 \\
    Enc. & 21.45 & $<$.001 & Lax: all/fullS p$<$.001 targetS/base p=.036; Tense: all/Stretch p$<$.001\\
    Resp. & 17.90 & $<$.001 & Lax: targetS/fullS,base p$<$.001, p=0.022 base/fullS p=0.001; targetS/fullS p$<$.001 base/fullS p=0.001\\
    \hline
  \end{tabular}
    \label{single_stats}
\centering
    {\parbox{6in}{
\footnotesize 
\centering
Tukey results are presented as: Higher value / Lower value\\
fullS = full stretch, targetS = target stretch
}}
\vspace{-5mm}
\end{table*}

\subsection{Experimental procedure}
We conducted an experiment to assess the objective (through word error rate) and subjective (through mean opinion scores) performance of French L1 English L2 listeners using our ``clarity mode'' compared to the baseline models (i.e., ``base'', ``stretch'', ``emphasis''). In an online experiment, participants chose to receive instructions in either English or French. They then provided demographic information on the language they first learned, their most commonly used daily language, their age and gender, and their self-rated English proficiency. Each participant was randomly assigned to start in either the single- or double-word trial.

The participants were shown the phrase with the target words removed, e.g., ``Write down the word X followed by the word X on the paper,'' and could play the audio only once. They then selected which words they heard and were told (in the double-word case) that it was possible to hear the same word twice. It was never the case that they would hear the same word, however, this instruction was added to encourage participants not to base their choice for the second word on what they believe they heard in the first word. The missing word was selected from a list of 4 words: for the single-word trial, 2 words were the tense/lax vowel minimal pair (e.g., beat, bit); the third word was a minimal pair word with another vowel (e.g., bat); and the final word was dissimilar from the other three choices (e.g., shop). The dissimilar choice functioned as an attention check. In the double word case, when the two words were a minimal pair, the selection for the remaining two words was as per the single-word trial; if the two words were not a minimal pair, the 4 choices were the two words and their minimal pairs. The order of the phrases was randomized for each participant. The experiment set up can be seen in Fig. \ref{ttsexp}.

Once the participants selected which word they heard, they responded to a questionnaire containing the MOS-X2 \cite{Lewis2018InvestigatingMR} naturalness (nMOS), intelligibility (iMOS), and prosody (pMOS) scores, as well as an intelligibility question for listening effort (eMOS) (``Please rate the degree of effort you had to make to understand the message'') from MOS-X \cite{Lewis2018InvestigatingMR}. The listening effort question was added to understand if there is a difference between being able to understand the phrase and how much effort was required to understand the phrase. They then responded to two questions from L2-directed speech research \cite{ROTHERMICH201922}: ``For an English second language speaker being spoken to with this voice, how respectful is the voice?'' (Resp.: condescending - respectful), ``For an English second language speaker being spoken to with this voice, how encouraging is the voice?'' (Enc: not encouraging - encouraging). These questions were to help us understand if slowing down or adding emphasis makes the TTS sound condescending, as can be the case with L2-directed speech \cite{AOKI2024101328, ROTHERMICH201922}. All scores were on a 10-point Likert scale.

\subsubsection{Participants}
We recruited N=56 participants (29F, age = 34.59$\pm$10.67) via Prolific. French was our target L1 population and the participant language demographics were as follows: daily language = French (40), English (14), French and English (2), Italian (1); English proficiency (1-5) = 5 (29), 4 (20), 3 (5), 2 (2); 71.4\% of participants chose to have the instructions in French. The study received internal ethics approval.

\subsection{Results}
\subsubsection{Single word}

\begin{figure}[]
  \centering
  \includegraphics[width=0.88\linewidth]{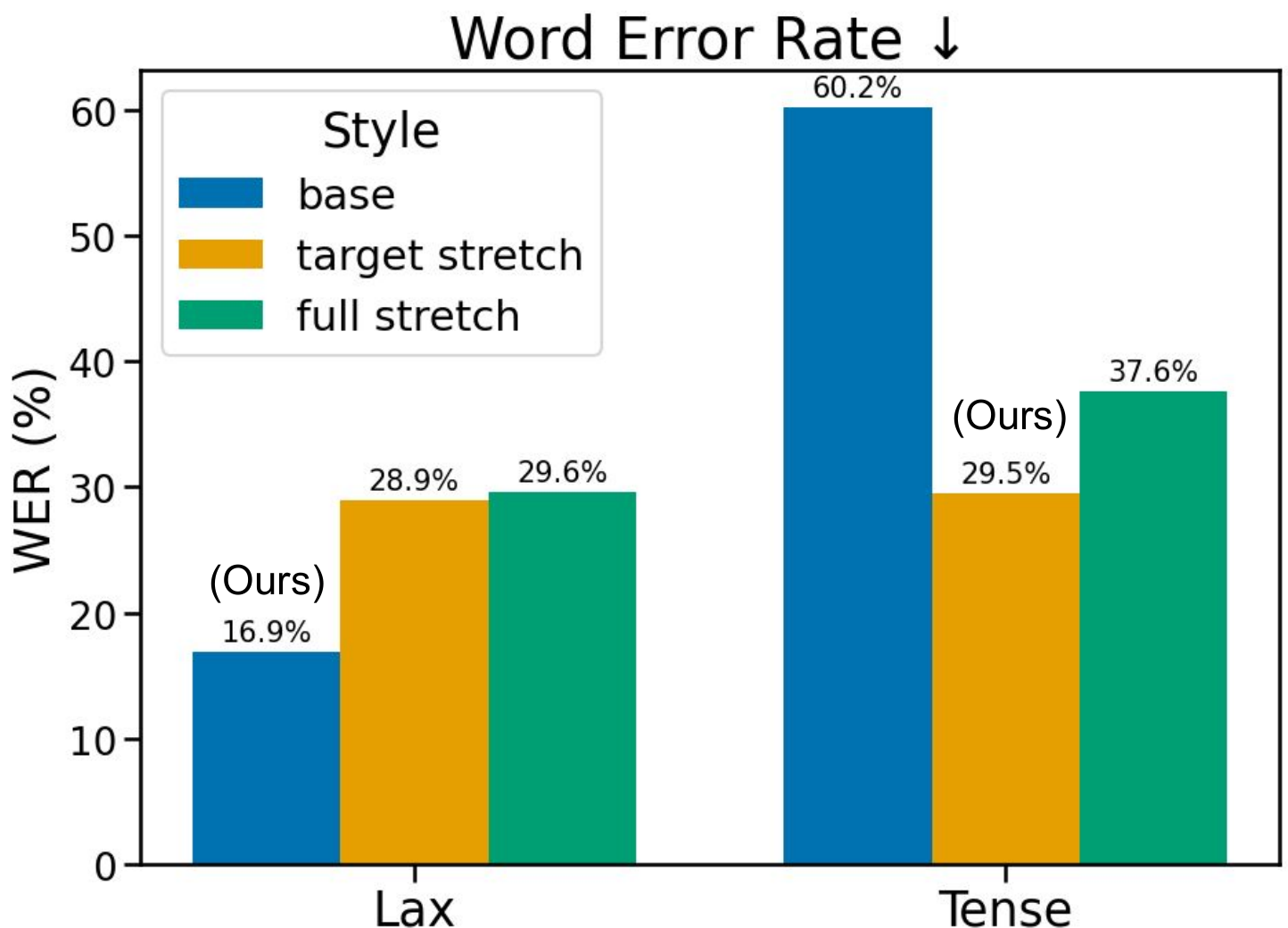}
  \caption{Single target word results: our clarity mode has the lowest WER for lax-base (left) and tense-target stretch  (right).}
  \label{wer-single-fig}
\vspace{-7mm}
\end{figure}

To confirm our previous findings, we expect improved performance in WER with lengthened tense vowels but decreased performance for the same treatment in lax vowels. We computed one-way ANOVAs (Type II) followed by a post-hoc Tukey HSD as well as word error rates (WER) as the proportion of incorrect target words for each type of target word treatment. In the single word case ``Clarity'' is the same as ``Base'' for lax vowel target words, and ``Emphasis'' for tense vowel target words. As such, we present the results in the three defined states existing for the single word case: ``base'' has the baseline duration on the target, ``target stretch'' is 1.6x the duration of the baseline on the target, and ``full stretch'' is 1.2x the duration on the entire phrase.

We observed that, through our ``clarity'' mode stretch applied to tense-vowel-containing words, we could overcome the bias towards lax vowels, i.e., the fact that participants were more likely to respond, for example, ``pill'' than ``peel,'' which we observed in our validation study. By lengthening the target tense-vowel-containing word, the WER on the baseline  (``base tense'') was reduced from 60.23\% to 29.48\% for ``target stretch tense'' (Table \ref{single_result}, Fig. \ref{wer-single-fig}). Importantly, we also confirmed that, indeed, stretching lax vowels, as is typical in L2-directed speech, results in more errors (28.90\% vs. 16.86\%). Lastly, the ``stretch'' TTS resulted in a slight reduction in WER over the baseline for tense vowels, yet not as much as for only emphasizing the target word. This suggests that a difference between the target word and its context was important for determining vowel perception \cite{tuttosi24_interspeech}. 

\begin{figure}[]
\vspace{-2mm}
  \centering
  \includegraphics[width=\linewidth]{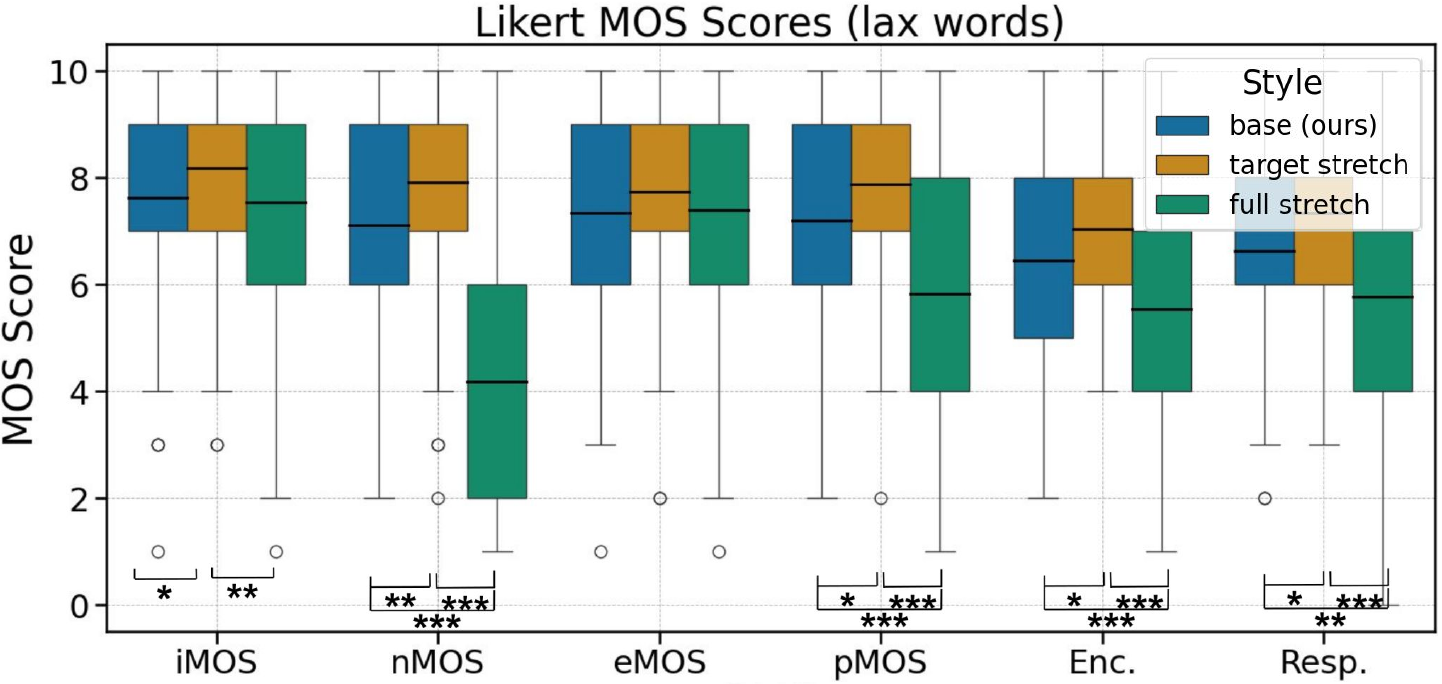}
  \caption{Single target word results: Likert MOS scores of lax vowel containing words for intelligibility (iMOS), naturalness (nMOS), effort (eMOS), prosody (pMOS), encouragement (Enc) and respect (Resp.) for base (ours), target stretch, and full stretch. Stretching the target word has significantly higher perceived intelligibility than both full stretch as the baseline, despite having lower objective comprehension performance. Both the baseline and target word stretch has significantly higher naturalness, prosody, encouragement and respectfulness over full stretch, and target word stretch over the baseline.}
  \label{mos-single-fig-lax}
\vspace{-5mm}
\end{figure}

\begin{figure}[]
\vspace{-2mm}
  \centering
  \includegraphics[width=\linewidth]{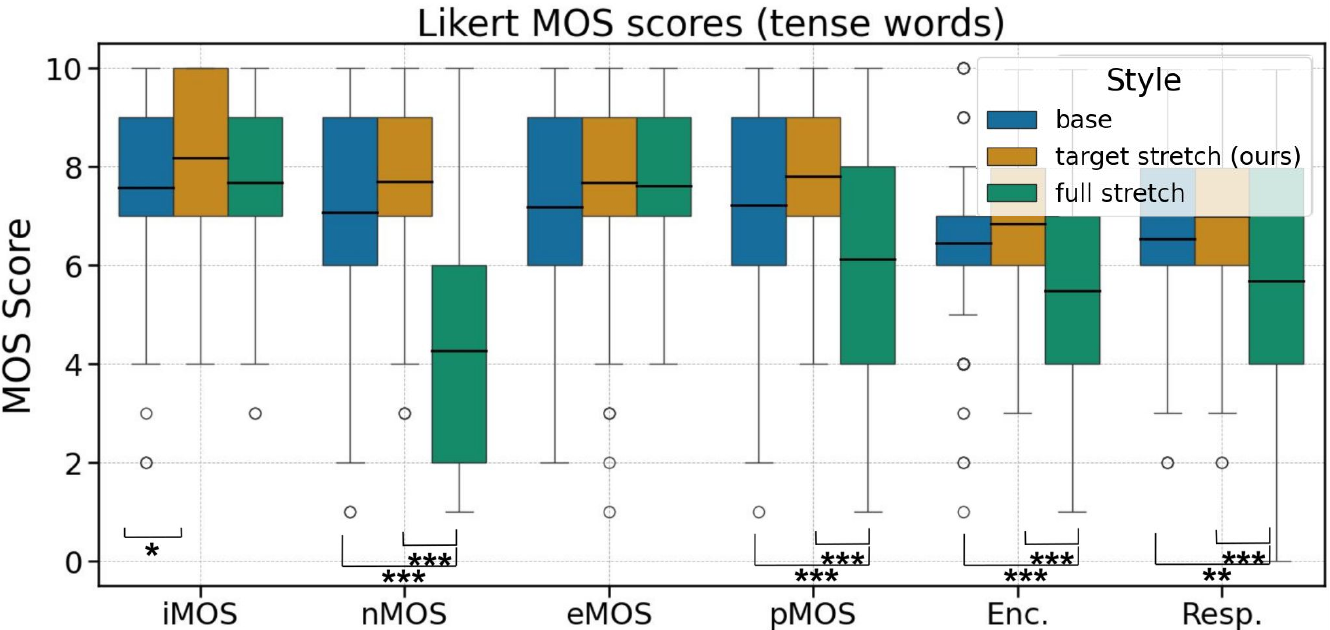}
  \caption{Single target word results: Likert MOS scores of tense vowel containing words for intelligibility (iMOS), naturalness (nMOS), effort (eMOS), prosody (pMOS), encouragement (Enc) and respect (Resp.) for base, target stretch (ours) and full stretch. Stretching the target word has significantly higher perceived intelligibility over the baseline. Both the baseline and stretching the target word have significantly higher naturalness, prosody, encouragement and respectfulness than the full stretch.}
  \label{mos-single-fig-tense}
\vspace{-5mm}
\end{figure}

Surprisingly, despite the objective improvements seen in the WER, the participants perceived a ``target stretch'' significantly ($\alpha = 0.05$) more intelligible for lax vowels (Tables \ref{single_result} and \ref{single_stats}, Figs. \ref{mos-single-fig-lax}, \ref{mos-single-fig-tense}). Moreover, ``target stretch'' also required less listening effort (although not significantly). Furthermore, although the ``full stretch'' TTS shows improvements in WER over the baseline for tense vowels, it was found to be significantly less respectful and encouraging in all cases. We also saw that the ``full stretch'' TTS was considered to be less natural and had worse prosody than the other TTS styles. This confirms our expectation that maintaining a natural (for L1 speakers) speech rate outside of difficult words can make the L2 listeners feel less patronized by the voice. Interestingly, the ``target stretch'' lax was rated as more natural and as having better prosody than ``base'' lax vowel, once again, despite have lower objective comprehension scores. These findings are particularly interesting as they suggest that L2 speakers may be overconfident in their ability to use the formant cues when the vowel is lengthened.

\subsubsection{Double word}

Once again, we computed WER as the proportion of correct target words for each TTS style. We also assessed differences between tense (tWER) and lax (lWER) vowels.

\begin{figure}[]
  \centering
  \includegraphics[width=\linewidth]{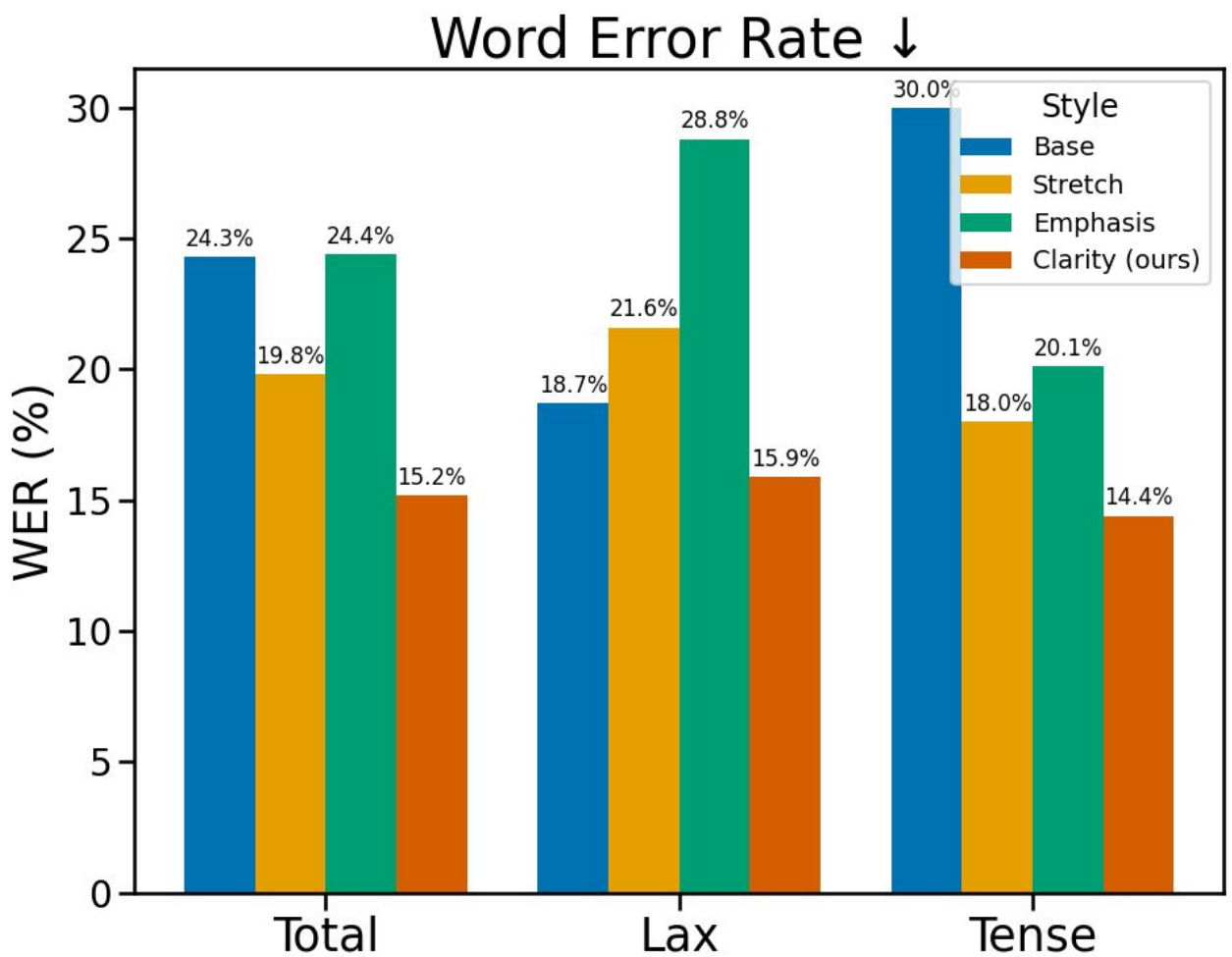}
  \caption{Double target word results: our clarity mode has the lowest total WER (left), lax WER (center) and tense WER (right).}
  \label{wer-double-fig}
\vspace{-5mm}
\end{figure}

We observed the lowest WER with our proposed ``clarity'' TTS, both overall and for lax vowels, while vastly improving the performance for tense vowels over the baseline (relative 50\% improvement) (Table \ref{double_wer}, Fig. \ref{wer-double-fig}). Again, the ``base'' TTS showed a bias towards lax vowels (tWER: 30.00\%, lWER: 18.66\%), that we were able to overcome with our ``clarity'' mode. Additionally, we see that simply emphasizing all difficult words that had a tense/lax minimal pair decreases performance in lax vowels. Lastly, we observed ``stretch'' having improved performance over both ``emphasis'' and ``base'' for tense vowels. Yet, as in the ``emphasis'' condition, we saw participants struggle to understand the short, lax vowels.  Although the tense vowel words between ``emphasis'' and ``clarity'' and lax vowel words between ``clarity'' and base had the same duration, ``clarity'' mode still had a lower WER when specifically comparing these words. This likely resulted from the phrases containing both a tense and a lax vowel, where the L2 participants could use duration differences between the two words to more easily differentiate the words. A more in-depth exploration of these differences remains a topic for future work. 

\begin{figure}[]
\vspace{-1mm}
  \centering
  \includegraphics[width=\linewidth]{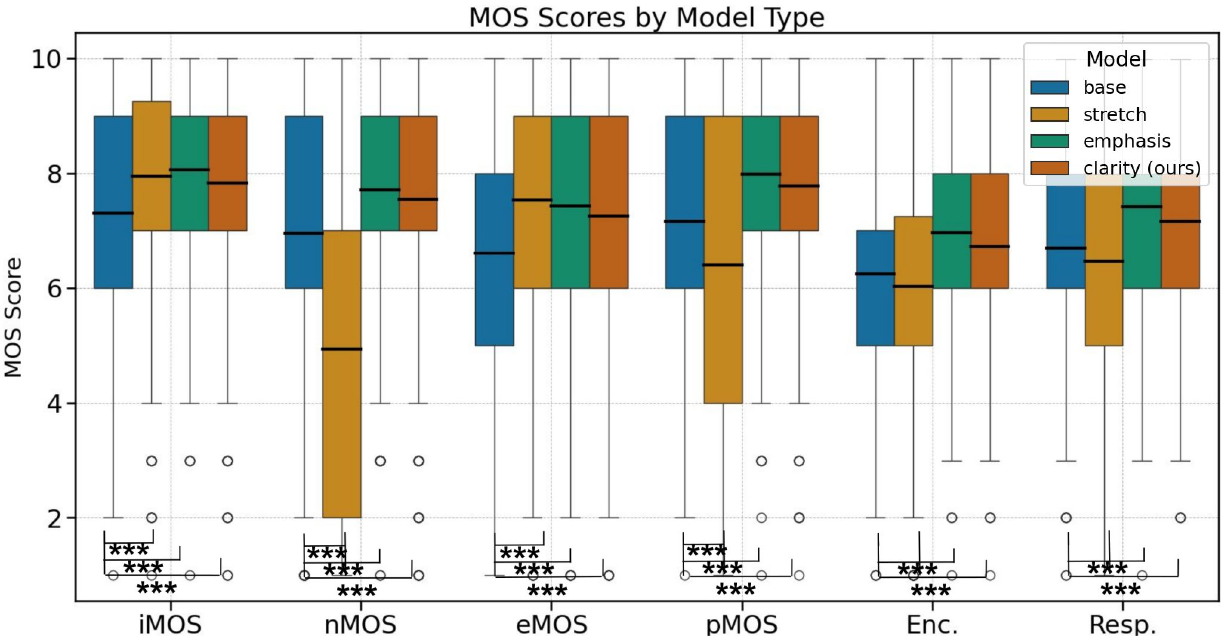}
  \caption{Double target word results: Likert MOS scores of words containing tense vowels for intelligibility (iMOS), naturalness (nMOS), effort (eMOS), prosody (pMOS), encouragement (Enc) and respect (Resp.). The baseline is perceived as  significantly less intelligible and requiring more listening effort than all of stretch, emphasis and clarity. Baseline and stretch are perceived as significantly less natural  with worse prosody than both emphasis and clarity, with stretch lower than baseline. Both baseline and stretch (speaking too fast or too slow) are perceived as significantly less encouraging and respectful.}
  \label{mos-double-fig}
\vspace{-5mm}
\end{figure}

We once again observed the fascinating result that the participants rate the ``emphasis'' TTS the highest in all categories (although in this case, the scores are not significantly higher than those for the ``clarity'' TTS) (Tables \ref{double_likert} and \ref{double_stats}, and Fig. \ref{mos-double-fig}), despite the WER being higher for this TTS than both ``clarity'' and ``stretch'' TTS styles. We observed that ``stretch'' is less natural and has poorer prosody than all other TTS styles, despite showing objective improvements in WER over ``emphasis'' and ``base''. Lastly, we found that L2 participants rated both speaking too fast (``base'') and too slow (``stretch'') as being less respectful and less encouraging.

\begin{table}[]
  \centering
  \caption{Double target word results: total word error rate, tense word error rate, and lax word error rate.}
  \vspace{-1mm}
  \begin{tabular}{| l | l | l | l | }
    \hline
    {\textbf{TTS Style}} & {\textbf{WER}} $\downarrow$ &  {\textbf{tWER}} $\downarrow$ &  {\textbf{lWER}} $\downarrow$\\
    \hline
    Base  & 24.30\% & 30.00\%  & 18.66\% \\
    Stretch & 19.82\% & 17.99\% & 21.65\%\\
    Emphasis &  24.44\% & 20.06\% & 28.82\% \\
    Clarity (Ours) & \textbf{15.15}\% & \textbf{14.38}\% & \textbf{15.92}\% \\
    \hline
  \end{tabular}
    \label{double_wer}
    \vspace{-3mm}
\end{table}

\begin{table*}[]
  \centering
  \caption{Double target word results: intelligibility, naturalness, effort, prosody, encouragement and respect.}
  \vspace{-3mm}
  \begin{tabular}{|  l |  l |  l |  l | l | l | l |}
    \hline
    {\textbf{TTS Style}} & {\textbf{iMOS}} $\uparrow$ & {\textbf{nMOS}} $\uparrow$ & {\textbf{eMOS}} $\uparrow$ & {\textbf{pMOS}} $\uparrow$ & {\textbf{Enc.}} $\uparrow$ & {\textbf{Resp.}} $\uparrow$\\
    \hline
    Base  &  7.30 & 6.95/*** & 6.60 & 7.16/*** & 6.25 & 6.70 \\
    Stretch & 7.94***/ & 4.93 & \textbf{7.53}***/ & 6.40 & 6.02 & 6.46\\
    Emphasis & \textbf{8.06}***/ & \textbf{7.71}***/*** & 7.43***/ & \textbf{7.98}***/*** & \textbf{6.97}***/*** & \textbf{7.42}***/***\\
    Clarity (Ours) & 7.83***/ & 7.54***/*** & 7.25***/ & 7.77***/*** & 6.72***/*** & 7.16***/***\\
    \hline
  \end{tabular}
    \label{double_likert}
    {\parbox{6in}{
\footnotesize 
\centering
Significance values are presented as: Significance compared to base TTS/Significance compared to stretch TTS
}}
\vspace{-3mm}
\end{table*}

\begin{table*}[]
  \centering
  \caption{Double target word results: ANOVA and Tukey test statistics on MOS Likert scores.}
  \vspace{-3mm}
  \begin{tabular}{|  l |  l |  l |  l |}
    \hline
    {\textbf{TTS Style}} & {\textbf{F-Statistic (3,2504)}} & {\textbf{P-Value}} & {\textbf{Significant Tukey Results}}\\
    \hline
    iMOS  &  20.15 & $<$.001 & all/Base p$<$.001 \\
    nMOS &  20.15 & $<$.001 & all/Stretch p$<$.001; Emphasis,Clarity/Base p$<$.001 \\
    eMOS & 24.15 & $<$.001 & all/Base p$<$.001 \\
    pMOS & 77.15 & $<$.001 & all/Stretch $<$.001; Emphasis,Clarity/Base p$<$.001 \\
    Enc. & 35.74 & $<$.001 & Emphasis,Clarity/Stretch,Base p$<$.001 \\
    Resp. & 35.87 & $<$.001 & Emphasis,Clarity/Stretch, Base p$<$.001 \\
    \hline
  \end{tabular}
    \label{double_stats}
\centering
    {\parbox{6in}{
\footnotesize 
\centering
Tukey results are presented as: Higher value/Lower value
}}
\vspace{-5mm}
\end{table*}

\subsubsection{Whisper ASR}

Speech synthesis studies often use human MOS scores but rely on ASR for transcription accuracy, as it correlates well with L1 intelligibility \cite{taylor21_interspeech}. In this section, we explore how ASR relates to L2 performance and whether it uses the same duration cues as L2 participants.

We used Whisper ASR \cite{radford2022robust}\footnote{v20231117, medium multilingual} with 72 phrases (generated in the same manner as for the human experiments) and calculated overall WER (WERt) and WER on only the target words. The phrases included those from the human study, and the additional phrases were constructed as in Sec. \ref{gen}. We also included the percentage of errors in the target word resulting from minimal tense/lax pair substitution (sub) and what percentage of these substitutions were lax substituted for a tense (t-sub) and tense substituted for a lax (l-sub) vowel. Additionally, we ensured that homophones with the target words were accepted as correct transcriptions.

\begin{figure}[]
  \centering
  \includegraphics[width=\linewidth]{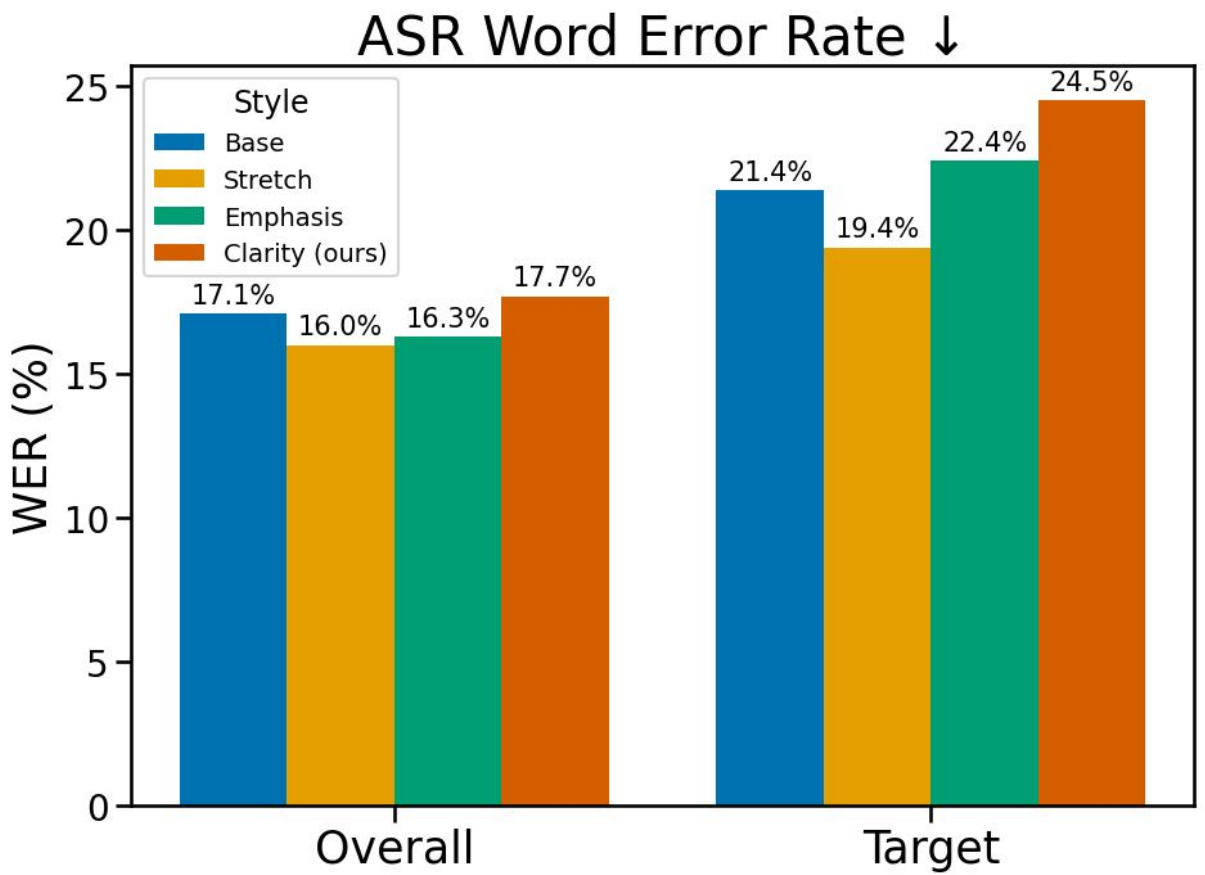}
  \caption{Whisper ASR results: We observe that ASR does not align with L2 perception (Fig. \ref{wer-double-fig}). Stretch had the lowest WER both for the whole phrase (left) and the target words (right).}
  \label{wer-asr-fig}
\vspace{-5mm}
\end{figure}

Similar to L2 speakers, we saw for the ``base'' TTS, Whisper struggled to predict the correct target words that lack context in the phrase (21.4\% vs an expected 5-10\% WER for this model), although the performance was slightly higher than that for the L2 speakers (Table \ref{whisper_wer}, Fig \ref{wer-asr-fig}). We did not see the same improvements in WER with the ``clarity'' TTS that we saw in the L2 participants. Instead, we saw an overall slowing down (``stretch'') of the TTS decreases the WER in the target words. Yet, we also saw that while the WER on the target words decreased with this slowing down, the proportion of errors in the target word stemming from the minimal pair substitutions was much higher for the ``base'' TTS (71.42\%, all other TTS $<$ 37\%), while the overall difference in WER both on the whole phrase and the target words was within 3\% for all TTS styles. This suggests that while slowing down slightly reduces the overall number of errors, the target words were being predicted as even further from the target, e.g. where ``peel'' was replaced with ``pill'' in the ``base'' TTS was replaced with ``peaked'' in the ``stretch'' TTS. Therefore, the ASR does not use the same duration mechanisms as humans when facing difficult predicting words.

\begin{table}[]
  \centering
  \caption{Whisper ASR results: overall word error rate, target word error rate, tense/lax substitutions, lax substituted for a tense, tense substituted for a lax}
    \vspace{-2mm}
  \footnotesize
  \begin{tabular}{| l | l | l | l | l | l | }
    \hline
    {\textbf{TTS Style}} & {\textbf{WERt}} $\downarrow$  & {\textbf{WER}} $\downarrow$ &  {\textbf{sub}} & {\textbf{t-sub}} &  {\textbf{l-sub}}\\
    \hline
    Base  & 17.10\% & 21.4\% & 71.42\% & 61.9\%  & 9.52\% \\
    Stretch & \textbf{15.98}\% & \textbf{19.38}\% & 36.83\% & 31.57\% & 5.26\%\\
    Emphasis & 16.26\% & 22.4\% & 31.81\% & 22.72\% & 9.09\% \\
    Clarity (Ours) & 17.68\% & 24.49\% & 29.16\%  & 29.16\% & 0\% \\
    \hline
  \end{tabular}
    \label{whisper_wer}
    \vspace{-5mm}
\end{table}

\section{Discussion and future work}

While we improved L2 comprehension of difficult words using duration cues for tense/lax minimal pairs, our focus remains limited to vowel length. Future work should explore clarity for other vowels via spectral cues (e.g., formants), and extend to consonants through manipulations like pauses and stress to enhance attention \cite{pausefocus}. Since we used a duration multiplier, clarity effects may depend on speech rate, which varies across speakers and could shift further with added expressivity such as emotion-laden speech. There may also be a minimum duration needed to perceive tense vowels, potentially unmet at faster rates. Understanding these limits—and how duration interacts with speech rate while preserving naturalness and respectfulness—remains an open area for investigation.

Further, through explorations of the results on the participant level in \cite{inprep}, we found that the duration mechanism was not clearly universal; Rather, it appears to be very strong in certain participants and weak or non-existent in others, which echoes work showing inter-individual differences in the weighting of L2 acoustic cues in speech perception \cite{schertz2015individual}. This suggests avenues for future work, which could harness pronunciation data to further customize TTS to individual listeners. Moreover, since our work used force choice to limit participant responses, a topic of future work is understanding how increasing clarity in a target word affects the rest of the phrase and how robust this duration mechanism is to broader linguistic contexts. Additionally, since our participants had a relatively high level of English proficiency, future work should explore ``L2 clarity TTS'' with participants of lower proficiency levels, which could explore the modification of words that are less likely to be in the vocabulary of L2 speakers (e.g., ``cooed'' vs. ``could''), using individual vocabularies as a factor predicting perception. Lastly, even more fine-grained control over a TTS \cite{tannander24_interspeech} could aid in uncovering other possible mechanisms to aid L2 perception, specifically for how formant control and duration control can be combined.

\section{Conclusions}
This study resulted in multiple interesting findings. First, we provided a ``clarity mode'' as an open-source addition to Matcha-TTS and confirmed that, by applying a stretch to tense vowels for difficult target words, L2 speakers' transcription errors for these words were reduced. Indeed, emphasizing all target words or simply slowing the whole phrase without consideration of the linguistic properties of the vowel reduced the transcription performance of L2 speakers (from French L1 backgrounds), who primarily use duration to determine the difference between English tense and lax vowels. Moreover, we confirm prior findings that slowing down an entire phrase can be seen as less respectful and encouraging to L2 listeners. 

Second, we found that our sample of L2 speakers are unaware they are using this duration mechanism. Through MOS scores, these L2 speakers indicated that they could more easily identify a stretched word, perhaps because they believed they could use the longer vowels (more stable formants) to transcribe a word more easily. This suggests that L2 listeners are limited in their ability to evaluate the effectiveness of adaptive TTS systems, and that objective metrics such as word error rate should serve as a gold standard in future work.

Lastly, we found that Whisper ASR does not use the same duration mechanisms as L2 speakers and, therefore, does not present an adequate replacement for determining the transcription accuracy of synthesized speech for these individuals. This suggests, again, that future development of adaptive TTS systems for L2 speakers must, at the point in time, rely more on data from real human listeners.

Overall, these results have important consequences for TTS assessments, especially for L2 speakers. We cannot simply rely on the same methods used for L1 speaker TTS assessments. Neither user self-rated intelligibility assessments nor automatic systems reflect the true accuracy of difficult vowel perception in L2 speakers, even for those with high proficiency as in our sample, and improved accuracy does not necessarily reflect positive perceptions of the voice (as for ``stretch''). Researchers must use both objective and subjective assessments with human participants to ensure they are building inclusive and accessible speech synthesis systems.

\section{Acknowledgements}
This work was supported by the Simon Fraser University FASS Breaking Barriers Interdisciplinary Incentive Grant, the Social Sciences and Humanities Research Council of Canada Grant (SSHRC Insight Grant 435–2019–1065), and the NSERC Discovery Grant (RGPIN-2024-06519). The authors thank Paul Maublanc for always being our first pilot French speaker, as well as the Rajan Family for their support.

\bibliographystyle{IEEEtran}
\bibliography{mybib}

\end{document}